\documentclass[aps,prl,10pt,twocolumn,
notitlepage,amsfonts,citeautoscript,superscriptaddress,showpacs]{revtex4-2}

\usepackage{amsmath}
\usepackage{amssymb}
\usepackage{amsfonts}
\usepackage{bm}

\usepackage{graphicx}   % load once
\usepackage[dvipsnames]{xcolor}
\usepackage{lipsum}
\usepackage{listings}
\usepackage{xfrac}
\usepackage{wrapfig}
\usepackage{mathrsfs}
\usepackage{latexsym}
\usepackage{url}
\usepackage{soul}
\usepackage{cancel}
\usepackage{braket}
\usepackage[caption=false]{subfig}
\definecolor{dkgreen}{rgb}{0,0.6,0}
\definecolor{gray}{rgb}{0.5,0.5,0.5}
\definecolor{mauve}{rgb}{0.58,0,0.82}
\usepackage[normalem]{ulem} % enables \sout
\usepackage[colorlinks=true,citecolor=blue,linkcolor=blue,urlcolor=blue]{hyperref}

\usepackage{cleveref}
\crefname{equation}{Eq.}{Eqs.}
\Crefname{equation}{Equation}{Equations}
\crefname{figure}{Fig.}{Figs.}
\Crefname{figure}{Figure}{Figures}
\crefname{section}{Sec.}{Secs.}
\Crefname{section}{Section}{Sections}
\crefname{appendix}{Appendix}{Apps.}
\Crefname{appendix}{Appendix}{Apps.}
\crefname{paragraph}{Sec.}{Secs.}
\crefname{table}{Table}{Tables}

\newcommand{\ketbra}[2]{\ket{#1}\bra{#2}}

\newcommand{\transpose}{\top}

\makeatletter
\def\@fnsymbol#1{\ifcase#1\or $*$\or $\mathsection$\or $\diamond$\or
\mathsection\or \mathparagraph\or $\star$\fi\relax}
\makeatother

\makeatletter
\def\maketitle{
\@author@finish
\title@column\titleblock@produce
\suppressfloats[t]}
\makeatother

\begin{document}
\title{Tensor-network representation of excitations in Josephson junction arrays}

\author{Emilio Rui}
\email{emilio.rui@alice-bob.com}
\affiliation{Alice and Bob, 53 Bd du Général Martial Valin, 75015 Paris, France}
\affiliation{Laboratoire de Physique de l'École Normale Supérieure,
École Normale Supérieure, Centre Automatique et Systèmes,
Mines Paris, Université PSL, CNRS, Inria, Paris, France}

\author{Joachim Cohen}
\thanks{These authors co-supervised the project.}
\affiliation{Alice and Bob, 53 Bd du Général Martial Valin, 75015 Paris, France}

\author{Alexandru Petrescu}
\thanks{These authors co-supervised the project.}
\affiliation{Laboratoire de Physique de l'École Normale Supérieure,
École Normale Supérieure, Centre Automatique et Systèmes,
Mines Paris, Université PSL, CNRS, Inria, Paris, France}

\date{\today}

\begin{abstract}
We present a nonperturbative tensor-network approach to the excitation spectra of superconducting circuits based on Josephson junction arrays. These arrays provide the large lumped inductances required for qubit designs, yet their intrinsically many-body nature is typically reduced to effective single-mode descriptions. Perturbative treatments attempt to include the collective array modes neglected in these approximations, but a fully nonperturbative analysis is challenging due to the many-body structure and the collective character of these modes. We overcome this difficulty using the DMRG-X algorithm, which extends tensor-network methods to excited states. Our key advance is a construction of trial states from the linearized mode structure, enabling direct computation of excitations, even in degenerate manifolds, which was previously inaccessible. Our results reveal significant deviations from, and allow us to improve upon, previous perturbative treatments in the regime of low array junction impedance.
\end{abstract}

\maketitle

\textit{Introduction.} Large, lumped linear inductors are critical components in the design of modern superconducting quantum circuits for quantum information processing. Inductive shunts of this kind are central to various architectures, including lumped-element LC resonators~\cite{Masluk2012}, the fluxonium qubit~\cite{Manucharyan2012}, protected qubits, such as the $0-\pi$ qubit~\cite{Gyenis2021}, or based on Cooper-pair pairing \cite{Smith2020Jan,Smith2022Apr}, and the asymmetrically threaded SQUID (ATS)~\cite{Lescanne2020-15,Ulysses}. These shunts are typically realized using arrays of Josephson junctions -- so-called superinductors -- that achieve impedances exceeding the resistance quantum~\cite{Masluk2012,Manucharyan2012,Bell2012Sep,Crescini2023Jun}.
Josephson junction arrays are as well central to the design of near–quantum-limited amplifiers \cite{HoEom2012, Macklin2015, Roch2012}.
%\er{maybe add some on the fact that the nonlinearity is suppressed with the number of junctions}

\begin{figure}[!t]
    \centering
    \includegraphics[width=.95\linewidth]{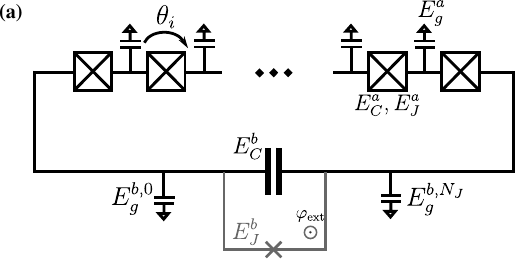}
    
    \vspace{1em} % Optional spacing between images
    
    \includegraphics[width=.95\linewidth]{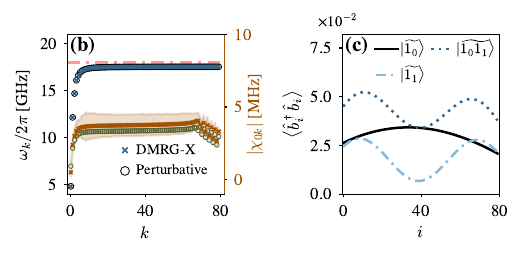}
    
    \caption{
        (a) Lumped LC resonator obtained through a capacitively shunted superinductance. The fluxonium qubit circuit is obtained by closing the loop with a small shunt ``black sheep'' junction (in grey).
       (b) DMRG-X–computed frequencies of the chain modes (blue crosses) compared to perturbation theory (black circles), and cross-Kerr nonlinearities $\chi_{0k}$ between the $k^{\textit{th}}$ mode and the fundamental mode (orange crosses) compared to perturbation theory (green circles). The shaded area indicates the uncertainty of the DMRG-X results, while the pink dot-dashed line marks the plasma frequency of the junctions in the array.
      (c) Chain-site occupation $\langle \hat{b}_i^\dagger \hat{b}_i \rangle$ as a function of site index $i$ for the single-excitation states corresponding to the fundamental mode $\ket{\widetilde{1_0}}$ (solid line) and the first mode $\ket{\widetilde{1_1}}$ (dot-dashed line), and for the two-excitation state $\ket{\widetilde{1_0 1_1}}$ (dotted line).
        }
    \label{fig:Fluxonium_circuit}
\end{figure}

In standard circuit quantization approaches, superinductors are modeled as a single lumped linear inductor, with the total phase drop across the array treated as a single dynamical variable (\cref{fig:Fluxonium_circuit}). In reality, however, each junction in the array has its own phase degree of freedom, giving rise to a set of collective chain or array modes~\cite{Ferguson2013,Viola2015,Warrington2024,Weil2015}. In typical implementations, the frequencies of these additional modes lie well above those of the fundamental modes relevant to the effective theory. Nonetheless, they could couple nontrivially to the low-energy sector, particularly in the presence of circuit imperfections~\cite{Ferguson2013}, nonlinearities~\cite{Viola2015}, or external parametric drives~\cite{Singh2025}. Such couplings could lead to spurious effects~\cite{Viola2015,Singh2025}, underscoring the need for accurate modeling.

In this Letter, we nonperturbatively study the excitation spectrum related to the internal-mode structure of Josephson-junction arrays, using tensor-network methods~\cite{White1992,DMRG_review,Vidal,Paeckel2019}. Previous work has shown that tensor networks can efficiently represent the low-lying spectra of fluxonium qubits~\cite{DiPaolo2021,Multi_Targeted}, while accurately accounting for the many-body structure of the Josephson junction array. 
Inspired by this, we focus here on the high frequency internal array modes, which, due to their high frequency and near-degeneracies around the characteristic plasma frequency, are numerically challenging.
To address this challenge, we employ the DMRG-X algorithm~\cite{DMRG-X}, a variant of the density-matrix renormalization group (DMRG)\cite{White1992,DMRG_review} designed to target individual excited states \cite{DMRG-X}. We thus exploit a recently pointed out correspondence between non-chaotic states in superconducting circuits and many-body localized systems \cite{Berke2022,Borner2024Aug}.

\begin{table}[t!]
    \centering
    \normalsize
    \setlength{\arrayrulewidth}{0.1mm}
    \begin{tabular}{|c|c|c|c|c|c|c|c|c|}
        \hline
        Set & $N_J$ & $E_{J}^{a}$ & $E_{C}^{a}$ & $E_{J}^{b}$ & $E_{C}^{b}$ & $E_{g}^{a}$ & $E_{g}^{b,0}$ & $E_{g}^{b,N_J}$ \\
        \hline
        1\cite{Masluk2012,Viola2015} & 80 & 84.3 & 0.483 & 0   & 6.07 &170 & 3.45 & 5.91 \\ % chain mode parameters
      2 \cite{Viola2015} & 43 & 26.0 & 1.24 & 8.93 & 3.60 & 194 & 4.80 &  4.80
        \\
        3\cite{DiPaolo2021} & 40 & -- & -- & 8.9 & 2.58 & 194 & 194 & 194 %Di Paolo parameter set
        \\
        % 4\cite{Viola2015} & 43 & 26.0 & 1.24 & 8.93 & 3.60 & 194 & 4.80 &  4.80 \\
        \hline
    \end{tabular}
    \caption{Device parameters for the three parameter sets. All energies are in GHz. The first parameter set describes the LC resonator studied in \cref{fig:Fluxonium_circuit}, the second in \cref{fig:Fluxonium}. The third parameter set describes \cref{fig:ng}. The missing terms are function of the array's junction impedance $z$ and their plasma frequency $\omega_p = 12.5$GHz}
    \label{tab:device-parameters}
\end{table}

DMRG-X enables the nonperturbative computation of excited states, provided a good initial trial state. It has previously been used as an effective diagonalization tool on top of black-box quantization approaches \cite{Rui_thesis,Putterman2025,diVincenzo2025,Gonzalez-Garcia2025Jun}. In this Letter, we extend this idea by developing a strategy to construct trial states from the normal modes of the linearized system, but perform no approximation at the level of the Hamiltonian. While we focused here on Josephson junction arrays -- particularly challenging for previous methods due to their degenerate spectrum -- this approach could be used for the nonperturbative study of systems usually treated by BBQ methods \cite{Nigg2012,epr}, without however expanding the Josephson nonlinearity \cite{Berke2022,Catelani2024,Borner2024Aug,diVincenzo2025,Gonzalez-Garcia2025Jun}.

We apply this framework to two experimentally relevant systems (\cref{fig:Fluxonium_circuit}). First, we consider a lumped-element LC resonator composed of a Josephson-junction array superinductance capacitively shunted~\cite{Masluk2012}, demonstrating that DMRG-X enables efficient and accurate resolution of the full spectrum of single-excitation chain modes. Second, we study the fluxonium qubit, focusing on the cross-Kerr nonlinearity between its fundamental mode and the most strongly coupled chain mode~\cite{Viola2015}. Our simulations reveal substantial quantitative deviations from previous perturbative predictions \cite{Viola2015, Singh2025}, highlighting the capability of the method to both validate and refine such approximations. Building on this, we propose improvements to perturbative treatments that significantly enhance their accuracy.
We eventually show that the method effectively captures charge-offset effects in the array~\cite{Matveev2002,Pop2010Aug,DiPaolo2021} by analyzing a fluxonium device whose superinductor is formed from more strongly nonlinear, high-impedance junctions.

% \er{Maybe addying also explicitely ng dependency}

% \subsection{Hamiltonian and guess}  
% \begin{itemize}
%     \item Introduction to Hamiltonian + reference to di Paolo for MPO representation + appendices
%     \item target state/linearized Hamiltonian
% \end{itemize}

%The DMRG-X algorithm was originally developed to target highly excited eigenstates with limited entanglement, especially in many-body localized (MBL) systems~\cite{DMRG-X}. In such contexts, trial states are typically chosen as product states in a localized basis (e.g., l-bit configurations).
%In this work, we generalize this strategy by constructing trial states as:
% \begin{align}
% \ket{(n_1)_{s_1} (n_2)_{s_2}\ldots} = \prod_{i=1}^k (A_{s_i}^\dagger)^{n_i} \ket{\mathrm{GS}},
% \end{align}
\textit{Model.} The system we study, depicted schematically in \cref{fig:Fluxonium_circuit}, is a superinductor of $N_J$ junctions, shunted either by a capacitor, realizing a lumped-element LC resonator~\cite{Masluk2012}, or by a small Josephson junction with a capacitance, realizing the fluxonium qubit~\cite{Manucharyan2012}.
The Hamiltonian describing this circuit~\cite{DiPaolo2021} is  
\begin{align} \label{eq:generic system}
\begin{split}
\hat{H} =& \sum_{i=1}^{N_J} \hat{H}_{0,i} + \sum_{i,j=1}^{N_J} \hbar g_{ij}\left(\hat{n}_i - n_{g,i}\right)\left(\hat{n}_j - n_{g,j}\right) \\
&+ U\!\left(\sum_{i=1}^{N_J} \hat{\theta}_i\right),
\end{split}
\end{align}
where $\hat{H}_{0,i}$ is a transmon-like Hamiltonian~\cite{Koch2007} describing the $i^\textit{th}$ junction in the array,  
\begin{align} \label{eq:localH}
\hat{H}_{0,i} = 4E_{C,i}(\hat{n}_i - n_{g,i})^2 - E_{J,i}\cos\hat{\theta}_i,
\end{align}
with $E_{J,i}$ the Josephson energy of the junction, $E_{C,i}$ its charging energy.
We explicitly include possible offset charges $n_{g,i}$ at each junction~\cite{DiPaolo2021}.
The second term in \cref{eq:generic system} accounts for long-range capacitive couplings set by the array layout and the shunt element \cite{SM}.  
The final contribution is the nonlinear potential of the black-sheep junction,  
\begin{align} \label{eq:potential}
U\!\left(\sum_i \hat{\theta}_i\right) = -E_{J}^b\cos\!\left(\sum_i \hat{\theta}_i + \varphi_{\mathrm{ext}}\right),
\end{align}
with $\varphi_{\mathrm{ext}}=2\pi\Phi_{\mathrm{ext}}/\Phi_0$ the dimensionless external flux and $\Phi_0$ the flux quantum.  
The lumped-element LC resonator is recovered in the limit $E_{J}^b=0$.

We represent the Hamiltonian \cref{eq:generic system} nonperturbatively as a matrix product operator (MPO) by expressing it in the eigenbasis of the local Hamiltonians \cref{eq:localH},
\begin{align} \label{eq:localHspec}
\hat{H}_{0,i} = \sum_{k=0}^\infty E_k^{(i)} \ketbra{\psi_k^{(i)}}{\psi_k^{(i)}} .
\end{align}
This representation, developed in Ref. \cite{DiPaolo2021} \cite{SM}, enables an efficient description of \cref{eq:generic system} while retaining the $n_{g,i}$-dependent charge effects~\cite{Koch2007}. However, in this basis the system’s eigenstates are not well captured by tensor-product states $\bigotimes_{i=1}^{N_J} \ket{\psi_{k_i}^{(i)}}$. In superinductors, the junctions are designed to be uniform, and this symmetry~\cite{Ferguson2013, Viola2015, Warrington2024} gives rise to collective excitations of the array. This poses a challenge for DMRG-X~\cite{DMRG-X}, which relies on tensor-product states of the local orbitals \cref{eq:localHspec} -- so-called \emph{l}-bit strings~\cite{diVincenzo2025, Gonzalez-Garcia2025Jun} -- as initial guesses.

To address this, we leverage the normal modes of the linearized system~\cite{Nigg2012,epr} as efficient starting points for constructing excited states with DMRG-X. Inspired by natural-orbital approaches~\cite{Debertolis2022,DeBertolis2025,Fernandez25}, we generalize the notion of a target state by promoting the $l$-bit string to a \emph{string of nonnegative occupation numbers ${l_i}$} associated with the normal modes defined below,
\begin{align} \label{eq:trial}
\ket{ {l_k} }_{\mathrm{trial}} = \prod_{k=1}^{N_J} (\hat{A}_{k}^\dagger)^{l_k} \ket{\Psi_0},
\end{align}
where $\ket{\Psi_0}$ is a reference state, typically the many-body ground state, and $A_{i}^\dagger$ denotes a finite-dimensional creation operator of the $i^\textit{th}$ normal mode.
To construct these operators, we start from ladder operators in the local eigenbasis,
\begin{align}\label{eq:b_local_basis}
\hat{b}_i = \sum_{k=1}^D \sqrt{k} \ketbra{\psi_{k-1}^{(i)}}{\psi_k^{(i)}}.
\end{align}
which reduce to \textit{bona fide} bosonic operators in the limit $D \to \infty$. The normal-mode operators $\hat{A}_i$ are then obtained via the symplectic transformation $S$ that diagonalizes the linearized Hamiltonian~\cite{Williamson1936,Colpa1978,Genoni2016,Serafini2017, SM},
\begin{align} \label{eq:symplectic}
\mathbf{\hat{A}} = S \mathbf{\hat{b}},
\end{align}
with $\mathbf{\hat{A}} = (\hat{A}_0, \hat{A}_0^\dagger, \hat{A}_1, \ldots)^\transpose$ and $\mathbf{\hat{b}} = (\hat{b}_0, \hat{b}_0^\dagger, \hat{b}_1, \ldots)^\transpose$.
While the coefficients in $S$ follow from a linear approximation, this affects only the construction of the trial state $\ket{{l_k}}$ in \cref{eq:trial}. The variational refinement in DMRG-X is then performed with the full MPO representation of the nonlinear Hamiltonian \cref{eq:generic system}, yielding the converged excited state $\ket{ \widetilde{{l_k}} }$.

\Cref{eq:generic system} encompasses a broad class of superconducting circuit Hamiltonians, beyond the specific superinductor configuration considered here \cite{Berke2022,Catelani2024}. The strategy we propose -- based on determining a set of natural excitations in a well controlled limit of the problem, and building trial states by constructing these excitations on top of a correlated reference state -- also applies to other architectures where a tensor-product trial state based on local orbitals may be inadequate. This could include fluxon lattices \cite{Petrescu2018}, transmon \cite{Berke2022,diVincenzo2025,Gonzalez-Garcia2025Jun}, and fluxonium arrays~\cite{diVincenzo2025}.

\begin{figure}[t!]
\centering
\includegraphics[width=.99\linewidth]{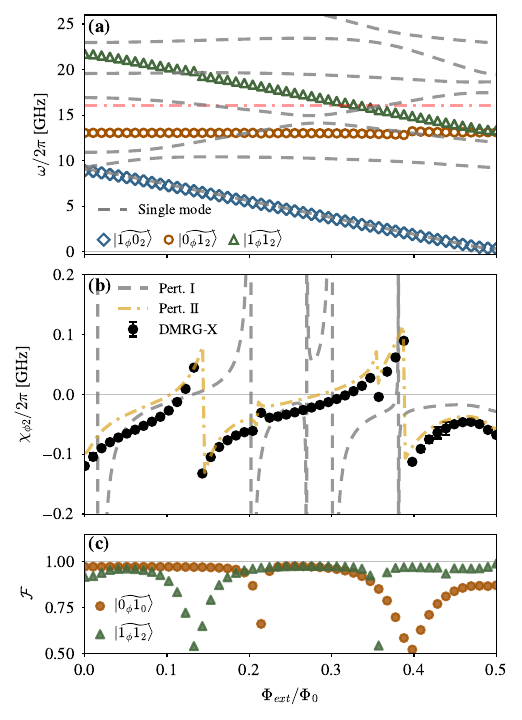}
\caption{DMRG-X applied to the Fluxonium qubit.
(a) Fluxonium energy spectrum as a function of the external flux $\varphi_{\mathrm{ext}}$ threading the circuit. The dashed gray lines show the effective-theory prediction from Ref.~\cite{DiPaolo2021}, while the dot-dashed pink line marks the plasma frequency of the junctions in the array. Blue markers denote the fluxonium state $\ket{\widetilde{1_\phi 0_2}}$ computed via DMRG, and orange and green markers indicate, respectively, the states $\ket{\widetilde{0_\phi 1_2}}$ and $\ket{\widetilde{1_\phi 1_2}}$ computed with DMRG-X.
(b) Cross-Kerr interaction $\chi_{\phi 2}$ between the fluxonium qubit and the first even chain mode. Dots represent the nonperturbative DMRG-X results. The dashed gray line shows the perturbative prediction from Ref.~\cite{Viola2015}, while the orange dot-dashed line includes corrections to the chain-mode frequency \cite{SM}.
(c) Fidelity $\mathcal{F} = \braket{\psi_{\mathrm{trial}} | \tilde{\psi}}$ between the DMRG-X–computed eigenstate $\ket{\tilde{\psi}}$ and the trial state $\ket{\psi_{\mathrm{trial}}}$.
}
\label{fig:Fluxonium}   
\end{figure}

\textit{Lumped LC resonator}.
We benchmark our method on a lumped-element LC resonator implemented as a capacitively shunted array of 80 Josephson junctions~\cite{Masluk2012}. The circuit is shown in black in \cref{fig:Fluxonium_circuit}a), with circuit parameters given in \cref{tab:device-parameters}, first row. This system is weakly anharmonic, and thus serves as an ideal testbed to validate our approach. 
We first probe the single-excitation manifold; to do so, we use as a reference state $\ket{\psi_{gs}}$ the ground state of the system using the standard DMRG technique \cite{DMRG_review, White1992}. Note that $\hat{A}_k \ket{\psi_{gs}} \approx 0$, with exact equality in the harmonic limit.  We consider trial states based upon \cref{eq:trial} corresponding to single excitations in the $k^\textit{th}$ mode, that is $l_k=1$ while $l_{k'}=0$ for all $k' \neq k$. 
% and then target excited states by applying approximate creation operators of the normal modes $\hat{A}_i^\dagger$ \cref{eq:symplectic}, yielding the trial state $\ket{1_{i,\text{trial}}} = \hat{A}_i^\dagger \ket{\psi_{gs}}$.

\Cref{fig:Fluxonium_circuit}b) compares the resulting excitation spectrum with predictions from  perturbation theory based on black-box quantization \cite{Nigg2012, epr, Petrescu2023,SM}. Our DMRG-X method converges to each of the $N_J$ single-excitation eigenstates and resolves their quasi-degenerate structure. The accuracy of each state, estimated from the energy variance, ${\sigma = (\braket{H^2} - \braket{H}^2)^{1/2}}$, remains below $1$ MHz. 
We achieve this small error bar by modifying the DMRG-X local optimization step \cite{DMRG-X} including shift-invert iterations \cite{diVincenzo2025, Yu2017, Baiardi2021, KrylovKit,Rui_in_prep}.

There is a slight renormalization of the mode frequencies compared to the linear normal modes, which becomes manifest in the slight discrepancy between the asymptotic, large $k$ value of the normal mode frequencies and the plasma frequency of the junctions. This correction is naturally captured by our nonperturbative approach, and is also accurately described by black box quantization based perturbation theory~\cite{epr, Nigg2012, SM}. Note that this contribution, with exceptions \cite{Weil2015}, has typically been neglected in previous analyzes of chain modes \cite{Masluk2012, Singh2025, Viola2015}.

Beyond the single-excitation sector, we access the two-excitation spectrum and compute cross-Kerr couplings between the fundamental mode and the other modes $\chi_{0j}$ as defined by the following Walsh-Hadamard transform~\cite{Berke2022} 
\begin{align}\label{eq:chi}
\chi_{ij} = \frac{1}{2} \left[ E(\ket{\widetilde{1_i 1_j}} - E(\ket{\widetilde{1_i 0_j}} - E(\ket{\widetilde{0_i 1_j}}\right],
\end{align}
with the understanding that all other modes have $l_k=0$, and all energies are measured relative to the ground state. These nonlinear couplings $\chi_{0j}$, shown in \cref{fig:Fluxonium_circuit}, are frequently used in experiments to measure the frequencies of the chain modes  \cite{Masluk2012, Weil2015}.

 The computation of these quantities is typically prohibitive for standard DMRG methods, which must resolve the full spectrum up to the target excitation and therefore incur a large computational overhead~\cite{DMRG_review, Multi_Targeted}. In addition, the relevant eigenstates are poorly represented in the tensor-product basis of local eigenstates $\ket{\psi_{k}^{(i)}}$ of the junction Hamiltonian $\hat{H}_{0,i}$ in \cref{eq:localH}, rendering a standard DMRG-X implementation unfeasible.
To illustrate this, we evaluate the junction occupations $\langle \hat{b}_i^\dagger \hat{b}_i \rangle$ for some single-excitation states. As shown in \cref{fig:Fluxonium_circuit}c), the resulting profiles follow the standing-wave patterns of the array~\cite{Masluk2012}, emphasizing their collective nature.%We further quantify this using the inverse participation ratios (IPR) of the DMRG-X eigenstates over tensor products of the local energy eigenbasis, while showing that the fidelity to the trial states \cref{eq:trial} remains high.% (for details, see \cref{ap:IPR}). %Low IPR values indicate that the DMRG-X eigenstates spread over many local basis states, confirming their non-product.%
%\footnote{TODO: Confirm and finalize the IPR definition and notation.}

\textit{Fluxonium qubit.}
We now extend our investigation to a regime of stronger nonlinearity by considering the fluxonium qubit~\cite{Manucharyan2012,Viola2015}. This circuit is described by the superinductor in \cref{eq:generic system} together with the potential energy \cref{eq:potential} of a single small Josephson junction -- the `black sheep' -- which introduces strong anharmonicity. Effective theories~\cite{Viola2015,Singh2025} show that the nonlinearity of the black sheep predominantly affects the fundamental mode of the system, which becomes the `fluxonium  mode', denoted $\phi = \sum_{i=1}^{N_J}\theta_i$, while higher-frequency array modes are relatively weakly perturbed by \cref{eq:potential}.
The residual coupling between the fundamental and the array modes, which can degrade the coherence time of the fluxonium qubit~\cite{Viola2015,Singh2025}, is well encapsulated by cross-Kerr interactions $\chi_{\phi i}$ of \cref{eq:chi}.

To compute $\chi_{\phi i}$, we extend the approach used for the LC resonator. Being strongly anharmonic, the excitations of the fluxonium mode are not well captured by the trial states \cref{eq:trial}. 
Exploiting the fact that the nonlinear fluxonium mode is the fundamental $0^\textit{th}$ mode \cite{Ferguson2013,Viola2015}, we first obtain the ground and first excited states of the full fluxonium Hamiltonian \cref{eq:generic system} using standard DMRG techniques~\cite{DiPaolo2021,Lee2003,Weiss2019}, yielding two reference states $\ket{\Psi_0}=\ket{\widetilde{0_\phi}}$ and $\ket{\widetilde{1_\phi}}$, where all the other modes are in their vacuum.
For example, to build the state $\ket{\widetilde{1_\phi 1_i}}$ with one excitation in both the fluxonium and the $i^\textit{th}$ chain mode, we construct $
{\ket{\psi_{\mathrm{trial}}} \equiv \ket{1_\phi 1_i} = \hat{A}_i^\dagger \ket{\widetilde{1_\phi 0_i}}}$, where $\ket{\widetilde{1_\phi 0_i}}$ is the first excited state of the fluxonium system with no added excitation in mode $i$.

We focus on the interaction between the fluxonium mode and the lowest even-parity array mode ($k=2$), which couples most strongly to the qubit~\cite{Viola2015}. \Cref{fig:Fluxonium}a) shows the circuit spectrum, for the parameters in \cref{tab:device-parameters}-2, as a function of the external flux $\varphi_{\mathrm{ext}}$. The single qubit excitation $\ket{\widetilde{1_\phi}}$ (blue circles) is obtained with standard DMRG as explained above, while the DMRG-X results for $\ket{\widetilde{0_\phi 1_i}}$ and $\ket{\widetilde{1_\phi 1_i}}$ are shown in orange and green, respectively. The corresponding cross-Kerr interaction $\chi_{\phi 2}$ is plotted in \cref{fig:Fluxonium}b).

The DMRG-X results (dots) show significant deviations from previous perturbative predictions (Pert. I)~\cite{Viola2015}, particularly near resonances and around the typical operating point $\Phi_{\mathrm{ext}}/\Phi_0=0.5$. Notably, our method converges reliably even close to degeneracies, where the quality of the trial states is expected to deteriorate. This is confirmed by the fidelity $\mathcal{F}=\braket{\psi_{\mathrm{trial}}|\tilde{\psi}}$ in \cref{fig:Fluxonium}c), which remains high across the flux range, aside from the expected drops at avoided crossings. The nonperturbative character of our approach thus makes it a powerful benchmark for perturbation theory. To this end, we refine previous treatments by incorporating the renormalization of chain-mode frequencies induced by array nonlinearities [Pert. II in \cref{fig:Fluxonium}b)], which substantially improves agreement \cite{SM}.

\begin{figure}
    \centering
    \includegraphics[width=\linewidth]{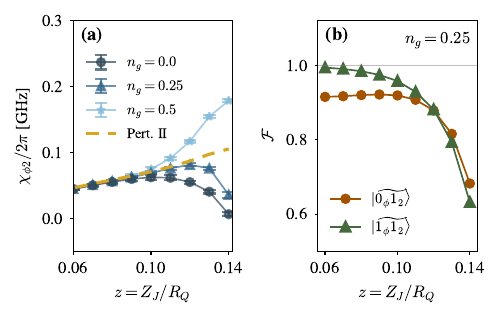}
    \caption{Charge dispersion of the cross-Kerr $\chi_{\phi2}$ in a fluxonium based on a $40$-junction superinductance \cite{DiPaolo2021} while increasing the array junction impedance $z = Z/R_Q$ at half flux quantum $\Phi_{ext}/\Phi_0 = 0.5$. The circuit parameters are listed in \cref{tab:device-parameters}. 
    (a) $\chi_{\phi2}$ between the fluxonium qubit and the second chain mode for a uniform gate charge in the array (solid lines) against the perturbative approach in the dashed line.
    (b) Fidelity $\mathcal{F} = \braket{\psi_\mathrm{trial}|\Tilde{\psi}}$ of the DMRG-X computed eigenstates $\ket{\Tilde{\psi}}$ and the trial states for $n_g = 0.25$.
    }
    \label{fig:ng}
\end{figure}

\textit{Offset-charge dependence}—The method proposed in this Letter retains the $2\pi$-periodic form of the Josephson potential and thus the compactness of the phase variables $\theta_i$~\cite{DiPaolo2021}. While alternative approaches expand the Josephson potentials in the Hamiltonian~\cite{Kaur2021}, we keep their full form \cite{SM}; the only approximations enter in the construction of trial states for DMRG-X. This allows us to capture offset-charge effects arising from the charges $n_{g,i}$ included in the Hamiltonian \cref{eq:generic system}, which, through the Aharonov–Casher effect~\cite{Pop2012,Friedman2002,Bell2016}, determine the phase of coherent quantum phase slips (CQPS) in the array~\cite{Koch2007,Manucharyan2012,Matveev2002,Mooij2005,Mooij2006,Hriscu2011,Rastelli2013}.

The fluxonium qubit provides a sensitive probe of coherent quantum phase slips (CQPS)~\cite{Manucharyan2012,manucharian_thesis}, as the black-sheep junction acts as a weak link in the array and CQPS lead to broadening of the qubit transition frequency. The CQPS rate depends exponentially on the reduced junction admittance $1/z$, making the effect increasingly pronounced for higher-impedance junctions. This regime has been previously investigated using tensor-network methods~\cite{DiPaolo2021}. Following the same strategy as in \cref{fig:Fluxonium}, we employ DMRG-X to study the offset-charge dependence of the cross-Kerr coupling $\chi_{\phi2}$ between the fluxonium and the $k=2$ chain mode in arrays with progressively larger impedance. 
%Since CQPS are dominated by the black-sheep junction, the fluxonium mode is the most strongly affected, allowing us to apply the same strategy as in \cref{fig:Fluxonium}. We therefore employ DMRG-X to study the offset-charge dependence of the cross-Kerr coupling $\chi_{02}$ between the fluxonium and the $k=2$ chain mode in arrays with progressively larger impedance.

In \cref{fig:ng}, we examine the effect of a uniform offset charge $n_{g,i} = n_g$ on the $z$-dependence of the cross-Kerr coupling $\chi_{\phi 2}$ between the fluxonium qubit biased at the half-flux point and the second collective mode of a $40$-junction array. As expected, CQPS become increasingly relevant at large $z$ \cite{Manucharyan2012}, inducing strong gate-charge sensitivity in the spectrum. In this high-impedance regime, the perturbative approach fails as it neglects the charge dispersion of the chain islands, with better agreement expected at $n_g=0.25$ where destructive interference suppresses coherent quantum phase slips \cite{Koch2007}. Nevertheless, our DMRG-X method converges across the entire range of impedance $z$ explored. 
We observe that the fidelity to the trial states decreases in the deep nonlinear regime, reflecting the breakdown of the harmonic approximation underlying their construction. Importantly, however, the method continues to converge even for large impedances $z>0.1$, demonstrating its robustness in strongly nonlinear regimes where perturbative approaches fail.

\textit{Conclusions}—We have proposed in this Letter an extension of the strategy to construct trial states for DMRG-X, enabling the targeting of collective excitations in Josephson junction arrays. This approach allows for fully nonperturbative computations without approximating the system Hamiltonian, while preserving the compact nature of the superconducting phase drop across each junction. We demonstrated that our method remains accurate even where perturbation theory fails, including the regime where quantum phase slips become prominent and the energy spectrum acquires a strong dependence on offset charges.

This approach opens the door to applying tensor network methods more broadly to superconducting circuits with many degrees of freedom. In particular, it provides quantitatively reliable access to chain mode dynamics in Josephson junction arrays, which may prove instrumental for future quantum information applications that seek to harness these modes as active resources.

\textit{Acknowledgment--} The authors thank Agustin Di Paolo, Serge Florens, Marco Genoni, Salim Miklass, Felix Rautschke, and Pierre Rouchon for stimulating discussions. Numerical simulations were performed on the CLEPS cluster at INRIA and relied on the Julia programming language \cite{bezanson2017julia} and the packages \texttt{ITensors.jl} \cite{ITensor} and \texttt{KrylovKit.jl} \cite{KrylovKit}.

\bibliographystyle{apsrev4-2}
\bibliography{Bibliography}

\end{document}

\title{Tensor-network representation of excitations in Josephson junction arrays}

\author{Emilio Rui}
\email{emilio.rui@alice-bob.com}
\affiliation{Alice and Bob, 53 Bd du Général Martial Valin, 75015 Paris, France}
\affiliation{Laboratoire de Physique de l'École Normale Supérieure,
École Normale Supérieure, Centre Automatique et Systèmes,
Mines Paris, Université PSL, CNRS, Inria, Paris, France}

\author{Joachim Cohen}
\thanks{These authors co-supervised the project.}
\affiliation{Alice and Bob, 53 Bd du Général Martial Valin, 75015 Paris, France}

\author{Alexandru Petrescu}
\thanks{These authors co-supervised the project.}
\affiliation{Laboratoire de Physique de l'École Normale Supérieure,
École Normale Supérieure, Centre Automatique et Systèmes,
Mines Paris, Université PSL, CNRS, Inria, Paris, France}

\date{\today}

\title{SUPPLEMENTARY MATERIAL\\
Tensor Network representation of excitations in Josephson junction arrays}
\maketitle

The Supplementary Material is structured as follows.
In \cref{app:Ham} we derive the Hamiltonian of the system under study, following Ref.~\cite{DiPaolo2021}.
In \cref{app:MPO} we introduce the tensor-network formalism and show how to efficiently represent the Hamiltonian in matrix-product form.
In \cref{app:target state} we derive the normal-mode bosonic operators from the linearized Hamiltonian.
In \cref{app:crossKerrLC} we incorporate the nonlinearity of the junctions in the array and present the perturbative computation of cross-Kerr interactions between the array modes.
In \cref{app:perturbation} we develop the perturbation theory for the fluxonium qubit, labeled “Pert. II” in the main text. While the approach is not fundamentally different from previous treatments \cite{Viola2015,Singh2025}, we formulate it in a way that facilitates numerical implementation, and validate it with an alternative parameter set in \cref{fig:pars1}.

\section{Superinductance Hamiltonian} \label{app:Ham}
% \er{this follows closely Ref.~\cite{DiPaolo2021}; we may decide whether it is worth keeping in full detail.}
In this section we derive the Hamiltonian of the system shown in \mainFig{1}. The derivation follows Ref.~\cite{DiPaolo2021} with the only difference that we explicitly perform a free-mode removal \cite{Ding2021}, which makes the analysis applicable also to nonuniform junction arrays. This refinement is marginal and is included here mainly for completeness. The same derivation reduces to a purely linear LC oscillator upon setting $E_{J}^b=0$. 
The Lagrangian describing the circuit\cite{Vool2017} in \mainFig{1} is 
% \begin{align}\label{eq:lagrangian_phi}
% \begin{aligned}
% L(\boldsymbol{\Phi}, \dot{\boldsymbol{\Phi}}) 
% &= \frac{1}{2}\sum_{i=0}^{N_J} C_{J}^a \left(\dot{\Phi}_i - \dot{\Phi}_{i-1}\right)^2 +  \frac{1}{2}\sum_{i=1}^{N_J-1} C_{g}^a \,\dot{\Phi}_i^2
%  \\
% &\quad + \frac{1}{2} C_{g,0}\,\dot{\Phi}_0^2
%    + \frac{1}{2} C_{g,N}\,\dot{\Phi}_{N_J}^2 \\
% &\quad + \sum_{i=1}^{N_J} E_{J,i}^a 
%     \cos\!\left[ \frac{2\pi}{\Phi_0} \left( \Phi_i - \Phi_{i-1} \right) \right] \\
% &\quad + E_{J}^b 
%     \cos\!\left[ \frac{2\pi}{\Phi_0} \left( \Phi_{N_J} - \Phi_0 + \Phi_{\mathrm{ext}} \right) \right],
% \end{aligned}
% \end{align}
%
\begin{align}\label{eq:lagrangian_phi}
\begin{aligned}
L(\mathbf{\Phi}, \dot{\mathbf{\Phi}}) 
&=   \frac{e^2}{4}\sum_{i=0}^{N_J} \frac{\left(\dot{\Phi}_i - \dot{\Phi}_{i-1}\right)^2 }{E_C^a} +  \frac{e^2}{4}\frac{\left(\dot{\Phi}_0 - \dot{\Phi}_{N_J}\right)^2 }{E_C^b}
 \\
&\quad +  \frac{e^2}{4}\left(\sum_{i=1}^{N_J-1}\frac{\,\dot{\Phi}_i^2}{E_g^a}  + \frac{\,\dot{\Phi}_0^2}{E_g^{b,0}} 
   +  \frac{ \,\dot{\Phi}_{N_J}^2}{E_g^{b,N_J}}\right) \\
&\quad + \sum_{i=1}^{N_J} E_{J,i}^a 
    \cos\!\left[ \frac{2\pi}{\Phi_0} \left( \Phi_i - \Phi_{i-1} \right) \right] \\
&\quad + E_{J}^b 
    \cos\!\left[ \frac{2\pi}{\Phi_0} \left( \Phi_{N_J} - \Phi_0 + \Phi_{\mathrm{ext}} \right) \right],
\end{aligned}
\end{align}
where we have defined ${\boldsymbol{\Phi} = (\Phi_0, \ldots, \Phi_{N_J})^\transpose}$ as the vector of node fluxes, with $\Phi_0 = h/2e$ the superconducting flux quantum.
$E_{J,i}^a$ are the Josephson energies of the large junctions in the array, and $E_{J}^b$ is the Josephson energy of the small black-sheep junction.
$E_C^a$ and $E_C^b$ denote the charging energies of the array junctions and the black-sheep junction, respectively, while $E_g^a$ corresponds to the capacitance to ground in the array.
Asymmetries in the capacitance to ground at the boundary islands, $E_g^{b,0}$ and $E_g^{b,N_J}$, are also included, accounting for possible asymmetric coupling to external circuitry \cite{Viola2015}.

We adopt the basis
$
{\boldsymbol{\Theta}_\Sigma = (\Sigma,\Theta_1,\ldots,\Theta_{N_J})^\transpose,}
$
where $\Theta_i = \Phi_i - \Phi_{i+1}$ denotes the branch flux across the $i^\textit{th}$ junction of the array, and 
$\Sigma = \sum_{i=0}^{N_J} \Phi_i$ is the collective sum coordinate. 
In this representation, the sum mode $\Sigma$ appears as a free mode; however, it remains capacitively coupled to the array degrees of freedom. 
Following Ref.~\cite{Ding2021}, we eliminate $\Sigma$ to obtain a reduced description of the system.

The capacitance matrix is transformed as ${{\mathbf{C}_{\Theta}} = (\mathbf{R}^T)^{-1} \mathbf{C}_\Phi \mathbf{R}^{-1}}$ where $\mathbf{R}$ is the coordinate transformation such that ${\boldsymbol{\Theta}_\Sigma =\mathbf{R}\mathbf{\Phi}}$, 
by representing it in block form
\begin{align}
\mathbf{C}_{\Theta} =
\begin{pmatrix}
\mathbf{C}_{\Sigma\Sigma} & \mathbf{C}_{\Sigma\theta} \\
\mathbf{C}_{\Sigma\theta}^\transpose & \mathbf{C}_{\theta\theta}
\end{pmatrix},
\end{align}
we aim to eliminate the off-diagonal blocks, where $C_{\theta\theta}$ is the $N_J\times N_J$ submatrix corresponding to the junction coordinates, and $C_{\Sigma\Sigma}$ is the $1\times 1$ block corresponding to the sum mode.  We now define the transformed coordinates
\begin{align}
    \boldsymbol{\Theta}_{\Sigma}' = W\,\boldsymbol{\Theta}_\Sigma 
    = (\Sigma',\Theta_1, \ldots, \Theta_{N_J} )^\transpose,
\end{align}
such that the mode $\Sigma'$ is decoupled from the dynamics. The transformation matrix $W$ that achieves this is \cite{Ferguson2013, Viola2015} 
\begin{align}
\left(W^\transpose\right)^{-1} =
\begin{pmatrix}
 1& \mathbf{0} \\
-\,\mathbf{C}_{\Sigma\theta}^\transpose \mathbf{C}_{\Sigma\Sigma}^{-1} & \mathbf{I}_{N_J}
\end{pmatrix}.
\end{align}
The reduced capacitance matrix $\mathbf{C}$, excluding the decoupled free mode $\Sigma'$, is then the bottom-right $N_J\times N_J$ block of
\begin{align}
    \mathbf{C}' = \left(W^\transpose\right)^{-1} \,\mathbf{C}_\Theta \, W^{-1} = 
    \begin{pmatrix}
    \mathbf{C}_{\Sigma'\Sigma'} & \mathbf{0} \\
    \mathbf{0}& \mathbf{C}
\end{pmatrix}.
\end{align}

By denoting $\mathbf{\Theta} = (\Theta_1, \ldots, \Theta_{N_J} )^\transpose$ the reduced variables (neglecting the uncoupled mode), the Lagrangian of the system is 
\begin{align} \label{eq:circuit_Lagrangian}
\begin{aligned}
L(\boldsymbol{\Theta}, \dot{\boldsymbol{\Theta}}) 
&= \frac{1}{2} \, \dot{\boldsymbol{\Theta}}^\transpose \mathbf{C}\, \dot{\boldsymbol{\Theta}} \\
&\quad + \sum_{i=1}^{N_J} E_{J,i}^a \cos\!\left( 2\pi\frac{\Theta_i}{\Phi_0} \right) \\
&\quad + E_{J}^b \cos\!\left[ \frac{2\pi}{\Phi_0} \left( \sum_{i=1}^{N_J} \Theta_i + \Phi_{\mathrm{ext}} \right) \right].
\end{aligned}
\end{align}

A Legendre transform yields the Hamiltonian
\begin{align}
\begin{split}
H =&  \frac{1}{2}\,\boldsymbol{q}_{\Theta}^\transpose \mathbf{C}^{-1} \boldsymbol{q}_{\Theta} \\
     &-\sum_{i=1}^{N_J} E_{J,i}^a \cos \theta_i 
     -E_{J}^b \cos \left(\sum_{i=1}^{N_J} \theta_i+\varphi_{\mathrm{ext}}\right),
     \end{split}
\end{align}
where $\theta_i = 2\pi\Theta_i/\Phi_0$ and $\varphi_{\mathrm{ext}} = 2\pi\Phi_{\mathrm{ext}}/\Phi_0$. By explicitly separating the diagonal part of the  kinetic term and including gate charges $n_{g,i}$ in the array \cite{DiPaolo2021}, the Hamiltonian (upon quantization) can be written as
\begin{align}
    \begin{split}
    \label{eq:circuit_Hamiltonian}
        \hat{H} =& \sum_{i=1}^{N_J} \hat{H}_{0_i}
        + \sum_{i\neq j}^{N_J} \hbar g_{ij} \,(\hat{n}_i-n_{g,i})(\hat{n}_j-n_{g,j})
        \\ &- E_{J}^b \cos \left(\sum_{i=1}^{N_J} \hat{\theta}_i+\varphi_{\mathrm{ext}}\right),
    \end{split}
\end{align}
where 
\begin{align}
\hat{H}_{0,i} &= 4 E_{C,i} (\hat{n}_i-n_{g,i})^2 - E_{J,i}^a \cos \hat{\theta}_i, \label{eq:H0i} \\
E_{C,i} &= \frac{e^2}{2}\,\big[ \mathbf{C}_{\Theta}^{-1} \big]_{ii}, \\
g_{ij}  &= 2 e^2 \,\big[ \mathbf{C}_{\Theta}^{-1} \big]_{ij}.
\end{align}
This, upon quantization, gives our starting point Hamiltonian. Below, we discuss how this can be represented as a matrix-product operator.
\section{MPO representation of JJ array Hamiltonian} \label{app:MPO}
In this appendix, we explicitly show how to efficiently represent the Hamiltonian of junction arrays in matrix-product operator (MPO) form.
We first give a brief overview of the matrix product state/operator framework, and then analyze the bond dimension of each term in the Hamiltonian. Our analysis demonstrates that, despite the presence of long-range and nonlinear couplings, the Hamiltonian admits an MPO representation whose bond dimension scales only polynomially with the system size.
Furthermore, by allowing for small truncation errors, the representation can be compressed to a constant bond dimension.

A generic quantum state of an $N$-body system with local basis $\{\ket{s_i}\}$ can be represented as a matrix-product state (MPS) \cite{Schollwck2005,Schollwck2011,Vidal, Bridgeman2017,Ors2014}
\begin{align}\label{eq:MPS}
|\psi\rangle=\sum_{s_i=0}^{d-1} \sum_{a_i=1}^D M_{a_1}^{(1), s_1} M_{a_1, a_2}^{(2), s_2} \ldots M_{a_{N-1}}^{(N), s_N}\left|s_1 s_2 \ldots s_N\right\rangle
\end{align}
where each probability amplitude is expressed as a product of matrices $M^{(i), s_i}$.
This representation requires to store $O(d N D^2)$ elements, where $d$ is the local Hilbert space dimension, and $D$ is what is called the bond dimension. The bond dimension scales with the entanglement of the quantum state \cite{DMRG_review,Vidal}, making tensor networks particularly efficient at compressing low-entanglement, localized quantum states.
Throughout this work we allowed for maximum bond dimensions of $D = 250$.
This formalism can be extended to represent operators in the form of MPOs
\begin{align} \label{eq:MPO}
\hat{O}=&\sum_{s_i, s_i'=0}^{d-1}\sum_{a_i=1}^{D} M^{(1),s_1, s_1'}_{a_{1}} 
M^{(2),s_2, s_2'}_{a_1,a_{2}}
\ldots\\ \ldots & M_{a_{N-1}}^{(N), s_N, s_N^{\prime}}\left|s_1^{\prime} s_2^{\prime} \ldots s_N^{\prime}\right\rangle\left\langle s_1 s_2 \ldots s_N\right|
\end{align}
by adding an extra set of indices $s_i'$.
The storage cost for an MPO scales as $O(d^2ND^2)$. As in the MPS case, efficiency depends on how the bond dimension $D$ grows with system size, which is associated again with the locality of the interactions described by $\hat{O}$.
The multidimensional tensor $M^{(i), s_i,s_i'}_{a_{i-1}a_{i}}$ denotes a matrix of size $D \times D$ whose entries are local operators acting on the $i^\textit{th}$ system of size ${d \times d}$ \cite{sum_MPO}

In our system, we work in the energy eigenbasis of the local Hamiltonian 
$\hat{H}_{0,i}$ introduced in \mainEq{2}. The number of quantum systems 
is thus given by the number of junctions in the array, $N = N_J$, while the 
local Hilbert-space dimension $d$ is determined by the number of eigenstates 
retained for each junction. To construct this basis, we diagonalize 
$\hat{H}_{0,i}$ in the local charge basis \cite{Koch2007} and keep the 
$d=8$ lowest-energy eigenstates,
\[
\{\,\ket{s_i} \equiv \ket{\psi_{s_i}^{(i)}} \;\big|\; s_i = 0,\ldots,d-1 \,\}.
\]

To represent \cref{eq:circuit_Hamiltonian} as an MPO, we use the fact that the sum 
of two MPOs is itself an MPO, with bond dimension equal to the sum of the 
individual bond dimensions. This allows us to analyze each term of the 
Hamiltonian independently.
The first term of \cref{eq:circuit_Hamiltonian} is a sum of on-site operators and can be represented as an MPO of bond dimension $D=2$ \cite{sum_MPO}
\begin{align}
\sum_{i=1}^{N_J} \hat{H}_{0,i} &=
\begin{pmatrix}
    \hat{H}_{0,1} & \mathbb{I}
\end{pmatrix} \otimes
\begin{pmatrix}
    \mathbb{I} & 0 \\
    \hat{H}_{0,2} & \mathbb{I}
\end{pmatrix}\otimes
\cdots
\otimes
\begin{pmatrix}
    \mathbb{I} \\ \hat{H}_{0,N_J}
\end{pmatrix}.
\end{align}

The second term of \cref{eq:circuit_Hamiltonian} represents an all-to-all two-body coupling. It can be expressed as an MPO of bond dimension $D = 2N$ \cite{sum_MPO} by factorizing
\begin{align}
\sum_{i,j=1}^{N_J} g_{ij} \, \hat{n}_i \hat{n}_j
&= \sum_{i=1}^{N_J} \hat{n}_i \left( \sum_{j=1}^{N_J} g_{ij} \, \hat{n}_j \right).
\end{align}
The term $\sum_j g_{ij} \, \hat{n}_j$ is itself a sum of local operators and can be represented with bond dimension $D=2$.  
Since the product of two MPOs has bond dimension equal to the product of their bond dimensions \cite{sum_MPO}, and each $\hat{n}_i$ can be represented as an MPO with $D=1$
\begin{align}
\hat{n}_i &=
\begin{pmatrix} \mathbb{I} \end{pmatrix}
\otimes \cdots \otimes
\begin{pmatrix} \hat{n}_i \end{pmatrix}
\otimes \cdots \otimes
\begin{pmatrix} \mathbb{I} \end{pmatrix}.
\end{align}
We obtain an overall bond dimension $D = 2N_J$ before compression. By applying standard MPO compression techniques~\cite{White2005,Parker2020}, this value can be further reduced. For all parameter choices considered, we find that after minor truncations—comparable to machine precision—the bond dimension of the charge-coupling term is $D = 5$. We emphasize that the foregoing analysis remains valid in the presence of charge offsets, since their inclusion simply amounts to the substitution $\hat{n}_i \to \hat{n}_i - n_{g,i}$ in the preceding expressions.

The last term of \cref{eq:circuit_Hamiltonian} is the cosine associated with the black-sheep junction, which also represents an all-to-all coupling. It can be expressed as
\begin{align}
\cos\!\left( \sum_{j=1}^{N_J} \hat{\theta}_j \right)
&= \frac{e^{\,\iu\sum_{j=1}^{N_J} \hat{\theta}_j} + e^{-\iu\sum_{j=1}^{N_J} \hat{\theta}_j}}{2}.
\end{align}
Since
\begin{align}
e^{\,\iu\sum_{j=1}^{N_J}  \hat{\theta}_j} &=
\begin{pmatrix} e^{\,\iu\hat{\theta}_1} \end{pmatrix}
\otimes \cdots \otimes
\begin{pmatrix} e^{\,\iu\hat{\theta}_j} \end{pmatrix}
\otimes \cdots \otimes
\begin{pmatrix} e^{\,\iu\hat{\theta}_{N_J}} \end{pmatrix},
\end{align}
is already a tensor product, it is directly an MPO with bond dimension $D=1$.  
Thus, the cosine—without expanding the exponential—is an MPO with $D=2$.  
This property remains valid for any cosine of a linear combination of local operators, such as those appearing in the BBQ formalism \cite{Rui_thesis,Nigg2012,epr}.

\section{MPO representation of normal mode bosonic operators} \label{app:target state}
In this section, we discuss how to obtain the bosonic operators of the array normal modes, and how to express them as MPOs.
The linearized Hamiltonian of \cref{eq:circuit_Hamiltonian}, neglecting the black-sheep junction, is
\begin{align}\label{eq:LC_linear} 
\begin{split}
\hat{H}_{\mathrm{lin}} &=
\sum_{i=1}^{N_J} \left( 4 E_{C,i} \hat{n}_i^2 + \frac{\tilde{E}_{J,i}^a}{2} \hat{\theta}_i^2 \right)
+ \sum_{i,j=1}^{N_J} g_{ij} \, \hat{n}_i \hat{n}_j ,\\
\end{split}
\end{align}
where we introduced the renormalized Josephson energy
\begin{align}\label{eq:Ejcorrection}
    \tilde{E}_{J,k}^{a} = e^{-\eta_k/4} E_{J,k}^{a} ,
\end{align}
with \(\eta_k\) a correction factor associated with the normal ordering of the cosine potential \cite{Petrescu2023} determined by the transcendental equation
\begin{align}
    e^{-\eta_k/4} \eta_k^2 = \frac{8 E_{C\,k}}{E_J^{\,k}} .
\end{align}

\Cref{eq:LC_linear} is quadratic, therefore diagonalizable via a symplectic transformation $S_{\xi\theta}$~\cite{Williamson1936,Colpa1978,Genoni2016,Serafini2017,Citeroni2025}.
We define $\mathbf{\hat{r}_\theta}$ and $\mathbf{\hat{r}_\xi}$ as the vectors of
local and normal-mode canonical variables, respectively:
\begin{subequations}\label{eq:vectors}
\begin{align}
\mathbf{\hat{r}_\theta} &= 
\big( \hat{\theta}_1, \ldots, \hat{\theta}_{N_J}, 
       \hat{n}_1, \ldots, \hat{n}_{N_J} \big)^{\mathsf{T}}, \tag{1a} \\
\mathbf{\hat{r}_\xi} &= 
\big( \hat{\xi}_0, \ldots, \hat{\xi}_{N_J-1}, 
       \hat{n}_{\xi,1}, \ldots, \hat{n}_{\xi,N_J-1} \big)^{\mathsf{T}}
= S_{\xi\theta}\,\mathbf{\hat{r}_\theta}. \tag{1b}
\end{align}
\end{subequations}
In terms of these variables, the quadratic Hamiltonian reads
\begin{align}
    \begin{split}
    H_\mathrm{lin} &= \tfrac{1}{2}\,\mathbf{\hat{r}_\xi}^{\mathsf{T}} D \mathbf{\hat{r}_\xi} 
    = \sum_{j=0}^{N_J-1} \tfrac{\omega_j}{2} \left(\hat{\xi}_j^2 + \hat{n}_{\xi,j}^2 \right) \\
    &= \sum_{j=0}^{N_J-1} \omega_j \left(\hat{A}^\dagger_j \hat{A}_j + \tfrac{1}{2} \right),
    \end{split}
\end{align}
where the normal-mode bosonic operators are defined as
\begin{align}
    \hat{A}_j = \tfrac{1}{\sqrt{2}} \left(\hat{\xi}_j + \iu \hat{n}_j\right).
\end{align}
By defining the local bosonic operators as
\begin{align} \label{eq:boson}
    \hat{b}_k = \frac{\hat{\theta}_k}{\sqrt{2\eta_k}} 
     + \iu \sqrt{\frac{\eta_k}{2}}\,\hat{n}_k,
\end{align}
we can collect the bosonic operators into vectors of normal modes and local modes,
\begin{subequations}\label{eq:boson_vectors}
\begin{align}
\mathbf{\hat{A}} &= 
(\hat{A}_1,\ldots,\hat{A}_{N_J},\hat{A}_1^\dagger,\ldots,\hat{A}_{N_J}^\dagger)^{\mathsf{T}}
= T_1 \mathbf{\hat{r}_\xi}, \tag{a} \\
\mathbf{\hat{b}} &= 
(\hat{b}_1,\ldots,\hat{b}_{N_J},\hat{b}_1^\dagger,\ldots,\hat{b}_{N_J}^\dagger)^{\mathsf{T}}
= T_2 \mathbf{\hat{r}_\theta}. \tag{b}
\end{align}
\end{subequations}

We obtain eventually the desired linear transformation between normal mode bosonic operators and local ones
\begin{equation}
    \mathbf{\hat{A}} = T_1^{-1} S_{\xi\theta} T_2 \mathbf{\hat{b}},
\end{equation}
where $T_1$ and $T_2$ denote the respective transformations between bosonic and canonical variables.

In order to express the normal mode bosonic operators $\hat{b}_i$ as ladder operators in the local eigenbasis used to represent the Hamiltonian \cref{app:MPO},
\begin{equation}
\hat{b}_i \approx \sum_{k=1}^{d-1} \sqrt{k},\ketbra{\psi_{k-1}^{(i)}}{\psi_{k}^{(i)}} .
\end{equation}
Since the normal-mode operators $\hat{A}_k$ are linear combinations of the local operators, they can also be written as MPOs with bond dimension $D_A=2$. Trial states are obtained by applying these MPOs to a reference state $\ket{\Psi_0}$. Applying an MPO to an MPS produces another MPS with bond dimension equal to the product of the two; hence, if the reference state $\ket{\Psi_0}$ is well represented by a low-bond-dimension MPS \cite{DiPaolo2021}, each created excitation increases the bond dimension only mildly, enabling the practical use of the DMRG-X algorithm.

\section{Cross-Kerr between chain modes in an superinductor} \label{app:crossKerrLC}
We now use the linearized analysis in \cref{app:target state} to compute cross-Kerr nonlinearities in a superinductor associated with the nonlinearity of the array using the black-box quantization approach \cite{Nigg2012,epr,Petrescu2023, Weil2015}, as shown in \mainFig{1}. The superinductor Hamiltonian \cref{eq:circuit_Hamiltonian} (with no black-sheep junction involved) can be separated into a linear part \cref{eq:LC_linear} and a nonlinear part 
\begin{align}
\begin{split}
\label{eq:HgenericNM}
    \hat{H} &= \hat{H}_{\mathrm{lin}} + \hat{H}_{\mathrm{nl}} \\ &= \sum_{i=1}^{N_J} \omega_i \, \hat{A}_i^\dagger \hat{A}_i  - \sum_{j=1}^{N_J} E_{J,j}^a \left(\cos\hat{\theta}_j + \frac{e^{-\frac{\eta_j}{4}}}{2}  \hat{\theta}_j ^2\right).
\end{split}
\end{align}
The nonlinearity generates therefore self-Kerr and cross-Kerr terms between the normal modes of the array annihilated by the operators $\hat{A}_i$. When expressing the phase variables as linear combinations of the normal-mode bosonic operators $\hat{\theta}_j = \sum_k u_{jk} \hat{A}_k + u_{jk}^*\hat{A_k}^\dagger$, the cross- and self-Kerr coupling constants from first-order rotating-wave approximation \cite{Petrescu2023} are
\begin{align}
\chi_{j k}^{(1)}=-\frac{1}{4} \sum_{m=1}^{N_J} |u_{mj}|^2 |u_{mk}|^2 E_{\mathrm{J}, m}^{\prime},
\end{align}
which is defined in terms of Josephson energies renormalized by factors arising from normal ordering
\begin{align}
    E_{\mathrm{J}, m}^{\prime} = e^{-\frac{1}{4}\sum_k |u_{mk}|^2} E_{J,m}^a.
\end{align} 

We find that including these factors is crucial for the convergence of the highest-frequency chain modes. The induced nonuniformity of the Josephson energies $E_{\mathrm{J}, m}^{\prime}$ lifts the degeneracy and localizes the highest frequency modes near the two boundaries. This localization has two consequences: first, DMRG-X fails to converge when trial states are constructed without accounting for it; second, the appearance of two distinct tails in the cross-Kerr spectrum of \mainFig{1-b} reflects the boundary localization of the modes, while their asymmetry arises from the unequal capacitances to ground at the two ends of the array as listed in \mainTab{1}.

Before concluding this section, we note that in certain cases the calculation can be made more precise by properly accounting for residual quadratic terms that appear in the normal-ordered expansion of the array nonlinearity. To see this, note that, due to the canonical transformation, the quadratic terms in \cref{eq:HgenericNM} do not fully cancel. They amount to a residual quadratic part of the Hamiltonian.
\begin{align}
\begin{split}
     &\hat{H}_\mathrm{lin,res} =\\ \sum_{j=1}^{N_J} &\frac{E_{J,j}^a}{2} \left( e^{-\frac{\eta_j}{4}} - e^{-\frac{1}{2}\sum_k |u_{jk}|^2}\right)  \left(\sum_k u_{jk} \hat{A}_k + u_{jk}^*\hat{A_k}^\dagger\right)^2.
     \end{split}
\end{align}
It is however possible to iteratively consider these linear contributions, by symplectically diagonalizing the quadratic Hamiltonian resulting from adding this residual part to the diagonal normal-mode Hamiltonian found above.
\begin{align}
    \tilde{H}_\mathrm{lin} = \sum_k \omega_k^{(j)} \hat{A_k}^\dagger\hat{A_k} + H_\mathrm{lin,res}.
\end{align}
For the parameter regimes considered, this step can be iterated and converges quickly to a set of normal-mode coefficients $u^\infty_{jk}$.
In practice, the corrections are minor—below $0.1\%$ in both mode frequencies and Kerr coefficients—and only marginally affect the quality of the trial states in DMRG-X simulations.

\begin{figure}[t!]
    \centering
\includegraphics[width=.99\linewidth]{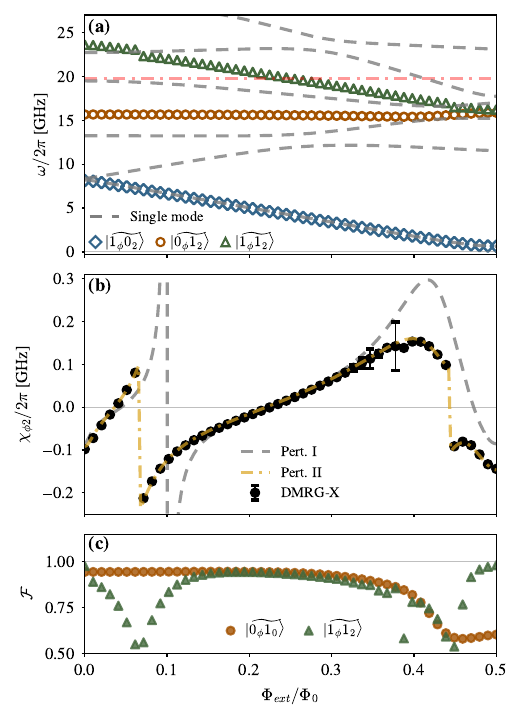}
   \caption{DMRG-X analysis of a fluxonium circuit with $N_J=95$ junctions in the superinductor (the circuit parameters are listed in \cref{tab:pars-appendix}). 
The analysis follows \mainFig{2}, but for a longer and less nonlinear array. 
In panel (b), the cross-Kerr coupling $\chi_{\phi 2}$ exhibits a marked deviation from the first-order perturbative result (Pert.~I), while showing improved agreement with the second-order perturbative approach (Pert.~II) due to the weaker nonlinearity.}

    \label{fig:pars1}
\end{figure}

\section{Cross-Kerr between the fluxonium and chain modes} \label{app:perturbation}
In this appendix we introduce the perturbative approach labeled “Pert. II” in the main text. 
Based on Ref.~\cite{Viola2015}, the method evaluates Kerr nonlinearities between the fluxonium mode and the array modes. Our treatment differs in that we perform numerically exact computations wherever possible and include frequency renormalizations from normal ordering of all bare modes. This leads to a clear improvement in precision when compared with exact numerics in the main text. We further validate the method in \cref{fig:pars1} using an alternative parameter set, listed in \cref{tab:pars-appendix}, corresponding to a fluxonium with a longer and less nonlinear array.

The nonlinearity caused by the array is significantly smaller than the one caused by the black sheep junction. Following \cite{Viola2015} we therefore focus on the cross-Kerr caused only by the black sheep junction.
Adding its term to the previous linearized Hamiltonian \cref{eq:LC_linear}, we have
\begin{align}\label{eq:Fluxonium_H_appendix} 
\begin{split}
\hat{H} &=
\hat{H}_{\mathrm{lin}} -E_{J}^b\cos\left(\sum_{i=1}^{N_J} \hat{\theta}_i + \varphi_{\mathrm{ext}}\right) \\
%&= \underline{r}^{\mathsf{T}} H \, \underline{r} -E_{Jb}\cos\left(\sum_{i=1}^{N_J} \hat{\theta}_i + \varphi_{ext}\right) ,x§x§
\end{split}
\end{align}
where we include the renormalization from normal-ordering as defined before in \cref{eq:Ejcorrection}. 
% , and the quadratic form on the second line is expressed in terms of 
% \begin{align}
% \underline{r} =
% \big( \hat{\theta}_1, \ldots, \hat{\theta}_{N_J},
%        \hat{n}_1, \ldots, \hat{n}_{N_J} \big)^{\mathsf{T}},
% \end{align}
% the vector of canonical variables. We now numerically symplecticly diagonalize the quadratic form $H$, namely we find a transformation matrix $S'$ such that $\underline{\tilde{r}} = S' \underline{r}$
% % \begin{align}
% % \underline{\tilde{r}} = S' \underline{r} =\big( \hat{\xi}_0, \ldots, \hat{\xi}_{N_{J-1}},
% %        \hat{n}_{{\xi}_0}, \ldots ,\hat{n}_{{\xi}_{N_{J-1}}} \big)^{\mathsf{T}},
% % \end{align}
% upon which the quadratic form becomes diagonal
% \begin{align}
%     \hat{H}_{lin} = \frac{1}{2} \tilde{ \underline{r}}^T D \tilde{ \underline{r}} 
% \end{align}
% where $D$ is a diagonal matrix of the normal mode frequency.    
Due to the block structure of the linear part, the symplectic transformation diagonalizing it $S_{\xi\theta}$ does not mix phase and charge, that is
\begin{align}
   S_{\xi\theta} =
    \begin{pmatrix}
        C & 0 \\
        0 & \left({C^\transpose}\right)^{-1}
    \end{pmatrix}.
\end{align}
This causes that normal mode phase variables $\xi_j$ are linear combinations of only the junction phase drops $\theta_i$. Importantly, the fundamental mode $\xi_0 \approx \sum_i \theta_i$, and therefore is the one mainly affected by the black-sheep nonlinearity. Following Ref. \cite{Viola2015} we want to enforce exactly the equality by the defining a ``fluxonium'' phase variable ${\phi = \sum_i \theta_i}$, so to move the whole nonlinearity to the fundamental mode. 
We define the canonical variables
\begin{align}
    \mathbf{\hat{r}}_\phi  = S_{\phi,\theta} \mathbf{\hat{r}} = \big( \hat{\phi},\hat{\xi}_1, \ldots, \hat{\xi}_{N_{J-1}},
       \hat{n}_{\phi},   \hat{n}_{\hat{\xi}_1},\ldots ,\hat{n}_{{\xi}_{N_{J-1}}} \big)^{\mathsf{T}},
\end{align}
where the symplectic matrix has the same structure as before, $S_{\phi,\theta} =  \mathrm{diag}(B,\left({B^T}\right)^{-1})$, with $B$ merely modifying the matrix $A$ introduced above by replacing all entries on its top row with ones 
\begin{align}
B &=
\begin{bmatrix}
1        & 1        & \cdots & 1        \\
C_{2,1}  & C_{2,2}  & \cdots & C_{2,N_J} \\
\vdots   & \vdots   & \ddots & \vdots   \\
C_{N_J,1} & C_{N_J,2} & \cdots & C_{N_J,N_J}.
\end{bmatrix}
\end{align}

\begin{table}[tp]
    \centering
    \normalsize
    \setlength{\arrayrulewidth}{0.1mm}
    \begin{tabular}{|c|c|c|c|c|c|c|c|c|}
        \hline
        Set & $N_J$ & $E_{J}^{a}$ & $E_{C}^{a}$ & $E_{J}^{b}$ & $E_{C}^{b}$ & $E_{g}^{a}$ & $E_{g}^{b,0}$ & $E_{g}^{b,N_J}$ \\
        \hline
        4\cite{Viola2015} & 95 & 48.3 & 1.01 & 10.2 & 4.78 & 484 & 11.9 & 11.9 %parameter set 2 VC
        \\
        \hline
    \end{tabular}
    \caption{Device parameters for the additional parameter set used in \cref{fig:pars1}. All energies are in GHz. }
    \label{tab:pars-appendix}
\end{table}

In terms of the transformed variables, the full system Hamiltonian reads  
\begin{align}
    \hat{H}
    = \hat{H}_\phi 
    + \tfrac{1}{2} \, \mathbf{\hat{r}_\xi}^{\mathsf{T}} H_{\xi} \, \mathbf{\hat{r}_\xi}
    + \sum_j g_{\phi j}^{(n)} \, \hat{n}_\phi \hat{n}_{\xi j} 
    + \sum_j g_{\phi j}^{(\varphi)} \, \hat{\phi} \hat{\xi}_j,
\end{align}
where $\mathbf{r_\xi}$ collects all normal-mode variables except the fluxonium ones $\hat{\phi}, \hat{n}_\phi$.  
Here, $H_{\xi}$ is an $2(N_J-1)\times2(N_J-1)$ matrix describing the residual couplings between chain modes;  
$g_{\phi j}^{(n)}$ denotes the capacitive coupling between the fluxonium and the $j^{\textit{th}}$ chain mode;  
and $g_{\phi j}^{(\varphi)}$ the corresponding flux coupling.  

The fluxonium Hamiltonian itself, obtained by isolating the $\phi$-mode charging and inductive terms from the transformed linearized Hamiltonian, is given by  
\begin{align}
     \hat{H}_\phi 
     = 4 E_C^\phi \, \hat{n}_\phi^2
     + \tfrac{E_L}{2} \, \hat{\phi}^2 
     - E_{Jb} \cos(\hat{\phi}+\varphi_{\mathrm{ext}}).
\end{align}
Note that $H_{\xi}$ is generally non-diagonal, as the change to $\mathbf{r}_\phi$ introduces off-diagonal terms determined by the circuit’s symmetries \cite{Viola2015,Ferguson2013}.

    We then symplectically diagonalize $H_\xi$, following the procedure outlined in \cref{app:target state}, to obtain dressed normal-mode bosonic operators for the array sector. This yields a star-coupling Hamiltonian in which the array modes couple only to the fluxonium mode,  
\begin{align} \label{eq:star coupling}
    \hat{H} = \hat{H}_\phi
            + \sum_k \left[ \omega_k \, \hat{a}^\dagger_k \hat{a}_k
            + \hat{n}_\phi \left( \tilde{g}_{\phi k} \, \hat{a}_k + \mathrm{h.c.} \right) \right],
\end{align}
where the effective couplings $\tilde{g}_{\phi k}$ now incorporate the normal-mode transformation.  
The Hamiltonian in \cref{eq:star coupling} thus describes a fluxonium mode linearly coupled to a set of harmonic oscillators (the array modes).  
Following \cite{Viola2015,Zhu2013}, one could apply perturbation theory to estimate the resulting cross-Kerr nonlinearities between the fluxonium and the array modes.  
Around resonances, an approach that agrees better with full numerics, however, is to restrict to a reduced subsystem consisting of the fluxonium and a single array mode \cite{Andersen24}, i.e., retaining only one term in the sum of \cref{eq:star coupling}.  
For the chosen $k^{\textit{th}}$ mode, the cross-Kerr $\chi_{\phi k}$ can then be computed by exact diagonalization \cite{epr} and careful state tracking \cite{Cohen}.

\bibliographystyle{apsrev4-2}
\bibliography{Bibliography}